\documentclass[a4paper,11pt]{article}
\pdfoutput=1 

\usepackage{jheppub} 
\usepackage[T1]{fontenc} 
\usepackage{wasysym}
\usepackage{tikz}
\usetikzlibrary{decorations.pathmorphing}
\tikzset{snake it/.style={decorate, decoration=snake}}

\usepackage{mathrsfs}

\newcommand*{\md}{{\cdot}}
\newcommand*{\e}{\varepsilon}

\title{\boldmath Scattering of spinning compact objects\\ from a worldline EFT}

\author{ Maor Ben-Shahar}
\affiliation{ Department of Physics and Astronomy, Uppsala University 
\\ Box 516, 75120 Uppsala, Sweden}

\emailAdd{benshahar.maor@physics.uu.se}

\preprint{UUITP-31/23}

\abstract{
We study the EFT of a spinning compact object and show that with appropriate gauge fixing, computations become amenable to worldline quantum field theory techniques. We use the resulting action to compute Compton and one-loop scattering amplitudes at fourth order in spin. By matching these amplitdes to solutions of the Teukolsky equations, we fix the values of Wilson coefficients appearing in the EFT such that it reproduces Kerr black hole scattering. We keep track of the spin supplementary condition throughout our computations and discuss alternative ways to ensure its preservation.
}

\begin{document}
\maketitle
\flushbottom
\newpage

\section{Introduction}
The recent discovery of gravitational waves~\cite{LIGOScientific:2016aoc,  LIGOScientific:2017vwq}, and expected improved sensitivity of future detectors~\cite{Punturo:2010zz,  LISA:2017pwj,  Reitze:2019iox}, call for precision theoretical predictions for astrophysical merger events. 
In the early ``inspiral'' phase of the two-body problem, many perturbative computations have been approached using effective fields theories~\cite{
Goldberger:2004jt,
Goldberger:2006bd,
Porto:2005ac,
Levi:2015msa,
Cheung:2018wkq,
Bern:2020buy,
Cheung:2020sdj,
Aoude:2020ygw}.
These often employ either the post-Newtonian (PN) approximation~\cite{
Goldberger:2004jt,
Goldberger:2006bd,
Porto:2005ac,
Levi:2015msa},
where one performs an expansion in Newton's constant $G$ and the velocity $v$, or the post-Minkowski (PM) approximation~\cite{
Aoude:2020ygw,
Bern:2020buy,
Bjerrum-Bohr:2023jau,
Cangemi:2022bew,
Chiodaroli:2021eug,
Chen:2021kxt,
Cheung:2018wkq,
Cheung:2020sdj,
Chung:2018kqs,
Damgaard:2019lfh},
where one expands only in Newton's constant.
In both regimes, actions based on a classical spinning compact object~\cite{Hanson:1974qy} have played a major role~\cite{
Porto:2008tb,
Porto:2008jj,
Porto:2010tr,
Porto:2010zg,
Levi:2010zu,
Levi:2011eq,
Porto:2012as,
Vaidya:2014kza,
Levi:2014gsa,
Levi:2014sba,
Levi:2015uxa,
Levi:2015ixa,
Levi:2016ofk,
Maia:2017gxn,
Maia:2017yok,
Siemonsen:2017yux,
Siemonsen:2019dsu,
Levi:2019kgk,
Levi:2020lfn,
Levi:2020kvb,
Levi:2020uwu,
Kim:2021rfj,
Cho:2021mqw,
Liu:2021zxr,
Cho:2022syn,
Kim:2022pou,
Mandal:2022nty,
Kim:2022bwv,
Mandal:2022ufb,
Levi:2022dqm,
Levi:2022rrq}.
In this paper we develop the Worldline Quantum Field theory (WQFT)~\cite{Mogull:2020sak,Jakobsen:2021zvh,Jakobsen:2021lvp} for EFT actions based on ref.~\cite{Hanson:1974qy}.
We will compute amplitudes beyond the spin-squared upper bound in current WQFT actions, based on supersymmetric worldline particles~\cite{Jakobsen:2021zvh, Jakobsen:2021lvp}, and use them to fix the Wilson coefficients in the spinning particle EFT by matching them to known observables for Kerr black holes~\cite{Bautista:2021wfy, Bautista:2022wjf}.

In the WQFT approach, computations are organized by introducing flat-space trajectories as background fields for the worldline degrees of freedom~\cite{Mogull:2020sak}. These then act and sources for worldline and metric fluctuations. 
The WQFT has been applied for the computations of Hamiltonians~\cite{Jakobsen:2022zsx} and radiation~\cite{Jakobsen:2022psy} with recent progress up to 4PM~\cite{Jakobsen:2022fcj,Jakobsen:2023hig,Jakobsen:2023ndj}. Using straight-line trajectories confines us to study scattering events, but in some cases, it was shown that scattering observables can be mapped to bound observables~\cite{Kalin:2019rwq,Kalin:2019inp}, or alternatively, it is possible to reconstruct two-body potentials from the amplitudes~\cite{Cheung:2018wkq,Neill:2013wsa,Vaidya:2014kza,Damour:2017zjx,Bjerrum-Bohr:2019kec}.
The WQFT and other worldline based approaches are usually manifestly classical, but potentials and observables with spin can actually be computed from quantum amplitudes~\cite{
Alessio:2022kwv,
Aoude:2020onz,
Aoude:2021oqj,
Aoude:2022trd,
Aoude:2023vdk,
Bern:2020uwk,
Bern:2022kto,
Bjerrum-Bohr:2023jau,
Cangemi:2022bew,
Chiodaroli:2021eug,
Chen:2021kxt,
Chen:2022clh,
Chung:2020rrz,
Chung:2018kqs,
Chung:2019duq,
Damgaard:2019lfh,
Guevara:2019fsj,
Guevara:2017csg,
Guevara:2018wpp,
Haddad:2021znf,
Haddad:2023ylx,
Heissenberg:2023uvo,
Kosmopoulos:2021zoq,
Maybee:2019jus,
Menezes:2022tcs,
Vines:2018gqi,
Vines:2017hyw}
after taking classical limits. This has led to interesting work on classical limits in general, and on the relationship between quantum and classical spin~\cite{Kosower:2018adc,Maybee:2019jus,Damour:2019lcq,Cangemi:2022abk}.

When working with EFTs it is not always obvious what principles should restrict Wilson coefficients in order to reproduce observables for specific astrophysical objects, for instance Kerr black holes.
At three points, amplitudes that reproduce the Kerr black hole's coupling to gravity exhibit a particularly simple structure~\cite{Vines:2017hyw,Guevara:2018wpp,Guevara:2019fsj}. There have been attempts to extrapolate these structures to higher orders using a so-called shift symmetry~\cite{Aoude:2022trd, Bern:2022kto, Haddad:2023ylx,  Aoude:2023vdk}, or gauge symmetries and power-counting constraints~\cite{Alessio:2023kgf,Cangemi:2022bew,Ochirov:2022nqz}.
At this stage, however, EFT computations have to be supplemented by matching to known Kerr observables or scattering amplitudes, such as those in refs.~\cite{Bautista:2021wfy,Bautista:2022wjf,Damgaard:2022jem,Siemonsen:2019dsu}. Higher-spin scattering amplitude computations from the spinning particle EFT have currently been done to spin cubed, confirming known results about Wilson coefficients for Kerr~\cite{Saketh:2022wap}. Additional constraints were found at fourth order in spin by computing the aligned-spin scattering angle~\cite{Siemonsen:2019dsu}, but some Wilson coefficients were still left unfixed. In the present paper, we compute the full Compton and one-loop scattering amplitudes at fourth order in spin, and are able fix all Wilson coefficients at this order by matching to refs.~\cite{Bautista:2021wfy,Bautista:2022wjf}. Beyond this order in spin, Kerr black holes already exhibit non-conservative effects~\cite{Bautista:2021wfy,Bautista:2022wjf}, which are usually modeled by the addition of extra degrees of freedom to the theory~\cite{Goldberger:2020fot, Saketh:2022xjb,Jones:2023ugm}.

A spin supplementary condition (SSC) is usually employed in worldline descriptions of rigid spinning bodies~\cite{Hanson:1974qy}, and it is often imposed by adding constraints with Lagrange multipliers directly to the worldline action. At low enough orders in spin the SSC and its conservation in time can be used to solve for the particle's momentum, which has been used to construct a Routhian description of the spinning body~\cite{Porto:2005ac}. This has been extended up to cubic order in spin, but may be difficult to extend to higher orders~\cite{Marsat:2014xea}. Alternatively, the second-class constraints imposing the SSC can be subsumed by a spin gauge symmetry~\cite{Steinhoff:2015ksa,Levi:2015msa,Vines:2016unv}, which when fixed should give various alternative but physically equivalent descriptions of the spinning particle.
For modeling spin-magnitude conserving processes it appears that the SSC is necessary~\cite{Vines:2017hyw, Chen:2021kxt,Chen:2022clh, Aoude:2022trd, Bern:2022kto,Bern:2023ity}, so we keep track of it throughout this work. 

This paper is organized as follows. We begin in section \ref{sec:action} by reviewing the construction of the spinning particle effective field theory, with emphasis on the gauge fixing. 
In particular we will remove the Lagrange multipliers which are an obstacle for perturbation theory, because they impose nonlinear constraints on the worldline fields. In sections \ref{sec:feyn} and \ref{sec:amps} we will present the Feynman rules and computations for the Compton and one-loop amplitudes. We present our conclusions in \ref{sec:con}. In appendix \ref{sec:A} we reproduce the full EFT action, and in \ref{sec:B} we explain the origin of the spin sector's cubic kinetic term from an action with a simpler kinetic term but a composite spin tensor. In appendix \ref{sec:c} we demonstrate an $R_\xi$-like gauge fixing which generalizes what we introduce in section \ref{sec:feyn}. Finally in appendix \ref{sec:d} we demonstrate that the constraints of the gauge-unfixed action are first class. Our full Compton and one-loop amplitudes are supplied in an ancillary file attached to this arXiv submission.

\section{The spinning particle effective field theory}
\label{sec:action}
We begin by recalling the action for the relativistic spinning particle~\cite{Hanson:1974qy,Porto:2005ac,Levi:2015msa}, which is a good model for spinning compact objects such as neutron stars or black holes.
Our strategy is to apply the machinery of the WQFT~\cite{Mogull:2020sak} to this action. This entails computing observables or scattering amplitudes directly from the path integral
\begin{align}
Z = \int \mathcal{D}[h]\mathcal{D}[W]e^{iS+iS_{EH}+S_{gf}} \ ,
\end{align}
where $h_{\mu\nu}$ is the metric fluctuation, $W$ represents the full set of worldline fluctuations $\{\pi^\mu,s^{\mu\nu},\lambda_{\mu\nu},z^\mu\}$ introduced in section \ref{sec:feyn}, $S$ is the action for the spinning particle(s), and the Einstein-Hilbert action $S_{EH}$ and gauge fixing terms $S_{gf}$ govern the metric self interactions. We work in de Donder gauge, so the latter two are
\begin{align}\label{eqn:eh}
S_{EH} = - \frac{2}{\kappa^2}\int d^4 x\sqrt{g} R 
 \ , \hspace{2cm}
 S_{gf} = \int d^4x \Big(\partial_{\nu}h^{\mu\nu} - \frac{1}{2}\partial_{\mu}h\Big)^2 \ ,
\end{align}
where $\kappa = \sqrt{32\pi G}$ is the gravitational coupling and $G$ is Newton's constant. 
The worldline action is written in terms of a Lagrangian density,
\begin{equation}
S = -\int d\tau \mathcal{L} \ ,
\end{equation}
which we will now construct.

\subsection{The worldline EFT}
Rotational degrees of freedom can be added to a worldline particle by introducing tetrad fields $e_\mu^I$ with Greek indices for the general spacetime manifold and capital Latin indices for the corotating, body-fixed frame.
We lower and raise indices with the spacetime metric $g_{\mu\nu}$ and the flat metric $\eta_{IJ}$ for the corotating frame, and use mostly-minus signature throughout. The tetrad $e_\mu^I$ can be decomposed further by writing  $e^\mu_I = e^\mu_a\Lambda^a_I$, where $e^a_\mu=e^a_\mu(x)$ now transforms from the spacetime manifold to a flat local frame, with metric $\eta_{ab}$, and $\Lambda_I^a=\Lambda_I^a(\tau)$ are Lorentz matrices defined on the worldline. 
We treat the fields $S_{ab}$ and $\Lambda_I^a$ as the fundamental spin degrees of freedom, but it is sometimes convenient to write them with curved indices using the frame fields $e_\mu^a$ and $e_a^\mu$. 
It will turn out later that in the scattering description the flat frame will be co-rotating with the particle in the asymptotic past, so the Lorentz matrices $\Lambda^a_I$ parametrize deviations of the particle's body-fixed frame at time $\tau$ from the body-fixed frame of the asymptotic particle. 

\begin{figure}[t]
	\centering
		\begin{tikzpicture}[scale=0.7]
		\draw[fill=blue!20,thick] (-2,0) rectangle (2,8);
		\draw[fill=blue!30,thick] (0,0) ellipse (2cm and .5cm);
		\draw[fill=blue!10,thick] (0,8) ellipse (2cm and .5cm);
		\draw[thick,draw opacity=0.6] (0,0) -- (0,8);
		\draw[thick,->,draw opacity=1] (0,8) -- (0,9);
		\node[] at (0,9.3) {$\tau$};
		\draw[scale=1, domain=0:2.4, smooth,draw opacity=0.2,thick, variable=\x, red] plot 
			({2*cos(1.3*deg(\x))}, {\x+ sin(deg(0.55))*2*sin(1.3*deg(\x))});
		\draw[scale=1, domain=2.4:4.85, smooth,thick, variable=\x, red] plot 
			({2*cos(1.3*deg(\x))}, {\x+ sin(deg(0.55))*2*sin(1.3*deg(\x))});
		\draw[scale=1, domain=4.85:7.3, smooth,thick,opacity=0.2, variable=\x, red] plot 
			({2*cos(1.3*deg(\x))}, {\x+ sin(deg(0.55))*2*sin(1.3*deg(\x))});
		\draw[scale=1,->, domain=7.3:9.3, smooth,thick, variable=\x, red] plot 
			({2*cos(1.3*deg(\x))}, {\x+ sin(deg(0.55))*2*sin(1.3*deg(\x))});
		\draw[thick,->,draw opacity=0.6] (0,0) -- (1.8,0);
		\draw[thick,->,draw opacity=0.6] (0,0) -- (-0.5,-0.4);
		\end{tikzpicture}
	\caption{
	The body-fixed frame tracks the orientation of a point (in red) on the rigid body as it rotates relative to the flat frame (in black). Only the spatial $SO(3)$ part of the $\Lambda_I^a$ fields are needed in order to achieve the rotation at a fixed time slice. The choice of point can be changed by a global $SO(3)$ little group transformation.
	}
	\label{fig:rotatingarm}
\end{figure}
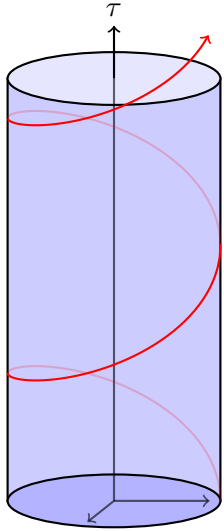

The Lorentz matrices carry six degrees of freedom, therefore in the Hamiltonian formalism they will be accompanied by six conjugate degrees of freedom, appearing through the antisymmetric spin tensor $S_{ab}$ (see Appendix \ref{sec:B} for an alternative construction). The spin tensor couples to the angular velocity
\begin{align}
\Omega^{ab} = \Lambda^{aI} \frac{D\Lambda^b_I}{d\tau} \ ,
\end{align}
via the kinetic term,
\begin{align}
\mathcal{L}_{kin} = \frac{1}{2} S_{ab}\Omega^{ab} \ .
\end{align}
As it stands, the spin sector contains six degrees of freedom in each of $S$ and $\Lambda$. Therefore one usually considers imposing the spin-supplementary condition~\cite{Hanson:1974qy} $S_{ab}p^b = 0$, as well as the conjugate constraint $\Lambda_{a0}= \hat{p}_a$, where we use the normalized momentum $\hat{p}^\mu\equiv p^\mu/\sqrt{p^2}$. 
This particular choice of constraints is called the covariant SSC.
Such constraints are motivated by examining the particle in its rest frame, where it is expected that a rigid body should have just three degrees of freedom in the spin sector, namely the boost components $\Lambda^a_{0}$ should not be dynamical~\cite{Steinhoff:2015ksa}, see Fig. \ref{fig:rotatingarm}.
Since it is challenging to apply perturbation theory to an action with Lagrange multipliers, it would be beneficial to find another formulation of the action such that it propagates the same degrees of freedom but is free from Lagrange multipliers.
The first step is to resolve the pair of second-class constraints $S_{ab}p^b=0$ and $\Lambda_{0a}=\hat{p}_a$ into a single constraint, at the cost of introducing a gauge symmetry. 
The introduction of the gauge symmetry was achieved in \cite{Levi:2015msa} by ``unfixing'' the gauge, such that the full minimally coupled worldline action is
\begin{align}\label{eqn:smin}
S = -\int d\tau \Big(
	p_\mu \dot{x}^\mu 
	+ \frac{1}{2} S_{ab}\Omega^{ab}
	+ \frac{1}{p^2}\frac{D p_\mu}{d\tau} S^{\mu\nu}p_\nu 
	-\frac{\ell}{2}(p^2-m^2)
	-\ell_ aS^{a b}\big(\hat{p}_b + \Lambda_{0b}\big) 
	\Big) ,
\end{align}
where $\ell$ and $\ell^\mu$ are Lagrange multipliers, and recall $\hat{p}^\mu$ is unit-normalized, $\hat{p}^\mu = p^\mu/\sqrt{p^2}$.
This action is not exactly in Hamiltonian form due to the acceleration term containing a derivative acting on the momentum. Several approaches have been implemented that do not have this acceleration term, for example by finite shifts of the worldline position~\cite{Vines:2016unv} or by solving the constraints for $p$~\cite{Porto:2005ac}.
But we will leave the acceleration term alone for the most part, as in the end it just contributes one extra term to the worldline propagators. 
Since the action \eqref{eqn:smin} was obtained by unfixing the covariant gauge, one can expect that the $\Lambda_0^a$ fields and the Lagrange multiplier $\ell^\mu$ will no longer be determined by the equations of motion. Therefore there should be some gauge symmetry in the action. The gauge symmetries can be guessed based on their form in flat space \cite{Steinhoff:2015ksa}, and indeed the following transformations leave the action invariant,
\begin{align}
\label{eq:gaugetransforms}
\delta S_{\mu\nu} =&
2 \hat{p}_{[\mu} S_{\nu]\alpha}\epsilon^\alpha \ , \nonumber \\
\delta \Lambda_I{}^\mu =& 2 \epsilon^{[\mu}\hat{p}^{\nu]}\Lambda_{I\nu} + 2 \epsilon^{[\mu}\Lambda_0^{\nu]}\Lambda_{I\nu} \ ,\\
\delta \ell^\mu =&-\frac{D{\epsilon}^\mu}{d\tau} +\ldots \nonumber \ .
\end{align}
The ellipses in the last gauge transformation include all additional terms needed to compensate for the variation of the action and constraint itself, we explain this below.
When Lagrange multipliers are undetermined, any variation of the action which results in something proportional to their associated constraint, in this case $S\md(\hat{p} + \Lambda_0)$, can be absorbed into a compensating transformation for the Lagrange multiplier itself. Therefore when checking gauge invariance one need only work on the constraint surface. Note that here the variation of the constraint itself also vanishes on the constraint surface, such that again it can be absorbed into the variation of the Lagrange multiplier again.
For the gauge transformations \eqref{eq:gaugetransforms} to be of physical significance the constraints in \eqref{eqn:smin} must be first class (see chapter three of ref.~\cite{Henneaux:1992ig}). We included a check of spin gauge independence of the amplitudes in Appendix \ref{sec:c}, and in appendix \ref{sec:d} we show that the constraint $S\md(\hat{p} + \Lambda_0)$ is preserved in time, implying it is indeed first class.

\paragraph{Conservation of spin magnitude:}
Before moving on to construct non-minimal couplings, it is worthwhile seeing the preservation of the spin magnitude from the gauge-invariant action.
We have an invariant spin tensor in the spatial part of the body-fixed frame, $S_{ij}=S_{ab}\Lambda^a_i\Lambda^b_j$, where $i,j$ run over $1,2,3$. This conserved charge comes from the global $SO(3)$ little group symmetry of the body-fixed frame (full $SO(1,3)$ is explicitly broken by the appearance of $\Lambda_{0}^a$ in the action). Since this charge is conserved, its square is also conserved, 
\begin{align}
S_{ij}S^{ij} =& \textrm{tr}\Big((\eta-\Lambda_0\Lambda_0)S(\eta-\Lambda_0\Lambda_0)S \Big)\ ,
\end{align}
and we can now recognize that in the covariant SSC ($\Lambda_0^\mu=p^\mu$ and $S^{\mu\nu}p_\nu=0$) the right-hand side is proportional to the magnitude of the spin vector $S^\mu = \frac{1}{2m}\epsilon^{\mu\nu\rho\sigma}p_\nu S_{\nu\rho}$.

\paragraph{Non-minimal couplings:}
Our next task is to add to \eqref{eqn:smin} additional spin and curvature dependent operators, which are necessary to model generic compact objects. 
Kerr black holes require some of these operators too, meaning they are not correctly modeled by the minimally coupled action. In section \ref{sec:amps} we will fix the new non-minimal couplings through fourth order in spin to match Kerr scattering.

It is clear that non-minimal couplings will be invariant under the gauge transformations \eqref{eq:gaugetransforms} if they are solely functions of the spin-vector, $S^\mu = \frac{1}{2m}\epsilon^{\mu\nu\rho\sigma}p_\nu S_{\nu\rho}$, or the spin tensor with transverse projectors $\tilde{S}_{\alpha\beta} = S_{\mu\nu}\mathcal{P}^\mu_\alpha\mathcal{P}^\nu_\beta$, where $\mathcal{P}^\mu_\alpha = \delta^\mu_\alpha - \hat{p}^\mu\hat{p}_\alpha$~\cite{Steinhoff:2015ksa}.
Since we will work in four dimensions, we will exclusively use the spin vector.
We first introduce the electric (parity even) and magnetic (parity odd) components of the Riemann tensor, 
\begin{align}
E_{\mu\nu}=& R_{\mu\rho\nu\sigma}p^\rho p^\sigma \ , \\
B_{\mu\nu}=& \frac{1}{2}R_{\alpha\beta \rho\mu}\epsilon^{\alpha\beta \gamma}{}_\nu p^\rho p_\gamma \ .
\end{align}
With these, all couplings linear in Riemann which respect the spin gauge symmetry are~\cite{Levi:2015msa},
\begin{equation}
\mathcal{O}(RS^{2n}) \sim D_{\mu_{2n}}\ldots D_{\mu_{3}}E_{\mu_1\mu_2}S^{\mu_1}\ldots S^{\mu_{2n}} \ ,
\end{equation}
for even powers in spin and 
\begin{equation}
\mathcal{O}(RS^{2n+1}) \sim D_{\mu_{2n+1}}\ldots D_{\mu_{3}}B_{\mu_1\mu_2}S^{\mu_1}\ldots S^{\mu_{2n+1}} \ ,
\end{equation}
for odd powers in spin.
These operators should be added inside the mass-shell constraint to maintain reparametrization invariance, so we add to our action
\begin{align}\label{eqn:SR}
\mathcal{L}_{R} = 
	\frac{\ell c_{ES^2}}{2m^2}E_{\mu\nu}S^\mu S^\nu
	- \frac{\ell c_{BS^3}}{6m^3}D_\mu B_{\nu\rho}S^\mu S^\nu S^\rho 
	-\frac{\ell c_{ES^4}}{24m^4}D_\mu D_\nu E_{\rho\sigma}S^\mu S^\nu S^\rho S^\sigma 
	 ,
\end{align}
which are all the couplings up to $S^4$ that are linear in Riemann. One should interpret $D_\mu D_\nu E_{\rho\sigma}= (D_\mu D_\nu R_{\rho\alpha\sigma\beta})p^\alpha p^\beta$ and similarly for derivatives of $B_{\mu\nu}$.
Aside from these, we will also add a single SSC-breaking operator,
\begin{align}\label{eqn:SK}
\mathcal{L}_{K} = 
	-\frac{\ell c_{RSK}}{2m^2} R_{\mu\nu\rho\sigma}S^{\mu\nu}K^\rho p^\sigma
	 \ ,
\end{align}
where $K^\mu = S^{\mu\nu}p_\nu$.
This operator would appear naturally in the action if we eliminate the acceleration term using the Mathisson–Papapetrou–Dixon equation,
\begin{align}\label{eqn:mpd}
\frac{Dp_\mu}{d\tau} = -\frac{1}{2}R_{\mu\nu\rho\sigma}\dot{x}^\nu S^{\rho\sigma} \ +\ldots,
\end{align}
in which case we would find \eqref{eqn:SK} with $c_{RSK}\to 1$, plus corrections at higher order in spin. In other words, if we worked without the SSC and the acceleration terms we would be able to obtain SSC-respecting results by fixing Wilson coefficients to definite values, see for example~\cite{Bern:2023ity} for similar observations. We will use this as a consistency check of our calculations, and we will not add other SSC breaking operators even though others could be written down.
At quadratic order in curvature and restricting to operators respecting the SSC we have (see e.g. refs.~\cite{Levi:2022rrq,Siemonsen:2019dsu}),
\begin{align}\label{eqn:SRR}
\mathcal{L}_{R^2} = 
	\frac{\ell c_{E^2S^4}}{ m^6}(E_{\mu\nu}S^\mu S^\nu)^2
	+\frac{\ell c'_{E^2S^4}}{m^6} S^2 S^\mu E_{\mu\nu}E^\nu{}_{\rho}S^\rho
	+\frac{\ell c''_{E^2S^4}}{m^6} S^4 E^2 + (E\leftrightarrow B)
	 .
\end{align}
In writing these operators we restricted to dimensionless Wilson coefficients which are independent of Newton's constant $G$, with the assumption that our compact object's characteristic size is not much greater than the Schwarzchild radius $2Gm$, and thus can not be used to soak up length scales without introducing factors of $G$ in the action. Note that for physical black holes, spin scales as $Gm^2$ so operators such as $S^4R^2/m^{4}$ contribute at the same order as $G^2S^2R^2$. 
But we will make comparisons to solutions of the Teukolsky equations at order $G^2$~\cite{Bautista:2021wfy,Bautista:2022wjf}, so we can stick to operators with no factors of $G$ in the action.
This concludes the construction of the action that we will use up to $S^4$, all terms are collected together in Appendix \ref{sec:A} for convenience.

\section{Gauge fixing and Feynman rules}
\label{sec:feyn}
In order to apply the WQFT techniques to the action in the previous section we need to study the flat-space equations of motion, and expand around their solutions. 
We gauge fix the action by setting (see appendix~\ref{sec:c} for a more general gauge fixing),
\begin{align}
\ell_\mu = \frac{1}{p}\frac{Dp_\mu}{d\tau} \ ,
\end{align}
which fixes our gauge to be the covariant SSC plus corrections at higher order in spin. We additionally fix reparametrization invariance by setting $\ell=\frac{1}{m} $. 
The minimally coupled action is then,
\begin{align}\label{eqn:smingaugefixed}
S =& -\int d\tau \Big(
	p_\mu \dot{x}^\mu 
	+ \frac{1}{2} S_{ab}\Omega^{ab}
	-\frac{1}{2m}(p^2-m^2)
	-\frac{1}{p}\frac{Dp_\mu}{d\tau}S^{\mu \nu} \Lambda_{\nu 0} 
	\Big) \ .
\end{align}
The flat space solutions have constant spin variables, $\dot{S}_{\mu\nu}=0$ and $\dot{\Lambda}^\mu_I=0$, and we have
\begin{align}
p_\mu = mv_\mu \ , \hspace{2cm} x^\mu= b^\mu + m v^\mu\tau \ , 
\end{align}
together with $v^2 = 1$. In addition, in the scattering scenario we will take our asymptotic states to obey the covariant SSC, meaning $\Lambda^\mu_0 = v^\mu$ and $S_{\mu\nu}v^\nu = 0$. 
We now use these solutions to the equations of motion as background fields, and add worldline fluctuations on top. 
This is achieved by the substitution,
\begin{align}
p_\mu &\to mv_\mu + \pi_\mu \ , \\
x^\mu &\to b^\mu + v^\mu \tau + z^\mu \ , \\
S_{ab} &\to S_{ab} + s_{ab} \ , \\
\Lambda^a_I &\to\Lambda_I^a + \lambda^{ab}\Lambda_{Ib} + \frac{1}{2}\lambda^{ab}\lambda_{bc}\Lambda^c_I +\ldots \ . 
\end{align}
From now on we interpret ${S}_{ab}$ and $\Lambda_I^a$ as the constant solutions to the equations of motion, and our dynamical fields are $\pi_\mu$, $z^\mu$, $s_{ab}$ and $\lambda_{ab}$. The $\lambda_{ab}$ matrices are antisymmetric, and are introduced through the exponential map of an infinitesimal rotation acting on the $\Lambda$ fields.
In principle we could have just written $\Lambda\to\Lambda+\delta$ but then $\delta$ must be constrained since $\Lambda+\delta$ must still be a Lorentz matrix. The antisymmetric infinitesimal rotation matrices trivialize these constraints and make it manifest that the $\Lambda$ sector contains six degrees of freedom. The downside of using these matrices is that we have an infinite set of vertices involving the $\lambda_{ab}$ fields. In practice only a small number of them will play a part in our classical perturbative computations.

\subsection{Propagators}
We can now expand the action up to quadratic order in the worldline perturbations and zeroth order in the gravitons in  order to compute the propagators. The quadratic part of the action is,
\begin{align}
S_{kin} =& -\int d\tau\Big(
\dot{z}^\mu\pi_\mu
- \frac{1}{2m}\pi^2
+ \frac{1}{2}S^{\mu\nu}\lambda_{\mu\rho} \dot{\lambda}^{\nu\rho}
- \frac{1}{2}\dot{\lambda}^{\mu\nu}s_{\mu\nu} \nonumber \\
&\hspace{6cm}
- \frac{1}{m} S^{\mu\nu}\dot{\pi}_\mu\lambda_{\nu\rho}v^\rho
- \frac{1}{m}\dot{\pi}^\mu s_{\mu\nu}v^\nu
\Big) \ ,
\end{align}
and we work in momentum/energy space so we need to insert the Fourier transforms,
\begin{align}
z^\mu(\tau) = \int \frac{d\omega}{2\pi} e^{i\omega t}z^\mu(\omega) \ , 
\end{align}
with similar expressions for the other worldline perturbations (here we work with incoming momenta and energies).
We are considering the propagation of one real particle, so when solving the free theory we pull out a factor of $\frac{1}{2}$ from the action and write the free partition function as $\textrm{exp}(\frac{i}{2}J\md \Delta\md J)$. Then the nontrivial two-point functions are,
\begin{align}\label{eqn:propagators}
	\langle z^\mu(-\omega) z^\nu(\omega)\rangle =& -i\frac{1}{m\omega^2}\eta^{\mu\nu} - \frac{1}{m^2 \omega} {S}^{\mu\nu}\ , \nonumber \\
	\langle p^\mu(-\omega) z^\nu(\omega)\rangle =& -\frac{1}{\omega}\eta^{\mu\nu}\ , \nonumber \\
	\langle s_{\mu\nu}(-\omega)s_{\rho\sigma}(\omega) \rangle =&  -\frac{2}{\omega} (
	\eta_{\nu[\sigma}{S}_{\rho]\mu}-\eta_{\mu[\sigma}{S}_{\rho]\nu}
	)\ , \\
	\langle s_{\mu\nu}(-\omega)\lambda_{\rho\sigma}(\omega)\rangle =& \frac{2}{\omega}\eta_{\mu[\rho}\eta_{\sigma]\nu}\ , \nonumber \\
	\langle \lambda^{\mu\nu}(-\omega)z_\rho(\omega)\rangle =& -\frac{2}{m\omega}v^{[\mu}\delta^{\nu]}_\rho \ . \nonumber
\end{align}
The retarded and advanced propagators are obtained by replacing $\frac{1}{\omega}\to \frac{1}{\omega\pm i0}$. Here we use the average of the two, meaning our background fields are the average of the far past/future background fields, see ref.~\cite{Mogull:2020sak} for more detail.
Note that only the last propagator introduces a mixing between the spin and position sectors, and it is the direct result of the acceleration and gauge-fixing terms. 
The propagator used for the gravitons is obtained by substituting $g_{\mu\nu}=\eta_{\mu\nu}+\kappa h_{\mu\nu}$ in equation \eqref{eqn:eh},
\begin{align}
\langle h_{\mu\nu}h_{\rho\sigma}\rangle =& 
\frac{i}{2(k^2+i0)}( 
		  \eta_{\mu \rho}\eta_{\sigma \nu}
		+ \eta_{\mu \sigma}\eta_{\rho \nu} 
		- \eta_{\mu \nu}\eta_{\rho \sigma}) \ .
\end{align}
We use the Feynman propagator which is suitable for the amplitude computations we do here.

\subsection{Vertices}
Simplified graviton self-interactions can be found in ref.~\cite{Kalin:2020mvi} for example, so we turn our attention to constructing vertives involving the worldline fields and gravitons.
We work in momentum space, so in the worldline action we substitute the Fourier transform
\begin{align}\label{eqn:hfourier}
h_{\mu\nu}(x) = \int\frac{d^4k}{(2\pi)^4} e^{ik\md(b+v\tau + z)}h_{\mu\nu}(k) \ ,
\end{align}
which when expanded in powers of $z$ introduces an infinite number of vertices coupling worldline perturbations to gravitons~\cite{Mogull:2020sak}. The extra $e^{ik\md b}$ factor will be dropped from the Feynman rules. 
We will also ignore the momentum integrals that come from the worldline action, and restore them at the end of the computation. If in the one-loop amplitude below we restore all momentum integrals that appear in the action we will be computing the Eikonal phase~\cite{Mogull:2020sak}, but to obtain the amplitude we will simply not integrate over the transferred momentum.
The $\tau$ integrals give us energy-conserving delta functions at each vertex, for example for a vertex with $m$ worldline perturbations and $n$ gravitons we have,
\begin{align}
\delta( \omega_1+\ldots +\omega_m + (k_1+\ldots +k_n)\md v ) \ ,
\end{align}
when all momenta and energies are incoming. We suppress these delta functions in the Feynman rules below.  

The simplest vertex we have has one graviton sourced by the background worldline trajectory. To construct it we work in momentum space and keep only terms linear in $h$ and at zeroth order in the worldline fluctuations. Supressing the coupling $\kappa$ we find
\begin{align}
\begin{tikzpicture}[scale=0.8,baseline={(0, -0.3cm)}]
	\draw[densely dotted] (0,0) -- (2,0);
	\draw[snake it,thick] (1,-1.3) -- (1,0);
\end{tikzpicture}
\ 
=
- \frac{i}{2}m (v \md h\md  v)
- \frac{1}{2} (v\md  h\md  S\md  k) -i\mathscr{L}_{\textrm{n.m.}}\Big|_{h}\ .
\end{align}
We represent the unperturbed worldline with a dotted line, and instead of including free indices in the vertex we used the field $h$ as an off-shell polarization tensor. This notation will become particularly convenient when dealing with the worldline perturbations below. 
We also included the non-minimal interactions through $\mathscr{L}_{\textrm{n.m.}}$, this is given by the Fourier transform of the non-minimal terms in $\mathcal{L}_{R}+\mathcal{L}_{K}$ processed in the same way as the contributions from the minimal action.
In this paper we only constructed non-minimal interactions up to $S^4$, but there is no barrier to working at higher orders once the relevant couplings are added.
At second order in the gravitons we have the vertex,
\begin{align}
\begin{tikzpicture}[scale=0.8,baseline={(0, -0.5cm)}]
	\draw[densely dotted] (0,0) -- (2,0);
	\draw[snake it,thick] (0.2,-1.3) -- (1,0);
	\draw[snake it,thick] (1.8,-1.3) -- (1,0);
\end{tikzpicture}
=&
\frac{im}{2}(v\md h_1\md h_2\md  v)
+ \frac{1}{4} (v\md h_1\md h_2\md S\md k_1)
+ \frac{1}{4} (v\md h_1\md S\md h_2\md k_1)
+ \frac{1}{8} {\rm tr}(h_2\md h_1\md S)v\md k_1\nonumber \\
&
+ (1\leftrightarrow 2) -i\mathscr{L}_{\textrm{n.m.}}\Big|_{h_1, h_2}\ ,
\end{align}
where now we introduced two placeholder polarizations $h_1$ and $h_2$, and now $\mathscr{L}_{\textrm{n.m.}}$ has contributions from $\mathcal{L}_{R} + \mathcal{L}_{R^2}+\mathcal{L}_{K}$. 
Finally, we will use one vertex that sources a worldline perturbation as well as a graviton,
\begin{align}
\label{eqn:Uvert}
\begin{tikzpicture}[scale=0.8,baseline={(0, -0.3cm)}]
	\draw[densely dotted] (0,0) -- (1,0);
	\draw[snake it,thick] (1,-1.3) -- (1,0);
	\draw[very thick] (1,0) -- (2,0);
\end{tikzpicture}
=&
\frac{m}{2}(v\md h\md v)(k\md z)
- i (\pi\md  h\md  v)
- \frac{i}{2} (v\md h\md S\md k)(z\md k)
- \frac{1}{2} (v\md h\md s\md k) \nonumber\\
&- \frac{i}{2}(z\md h\md S\md k)\omega+ \frac{1}{2}(v\md h\md v)(v\md \lambda\md  S\md k)
- \frac{1}{2}(v\md h\md v)(v\md s\md k) -i\mathscr{L}_{\textrm{n.m.}}\Big|_{h, W}\ ,
\end{align}
where again we abuse notation slightly by using the worldline fields $\{z^\mu, \lambda^{\mu\nu}, s_{\mu\nu}, \pi_\mu\}=W$ in the same way as we used $h$ above.
Notice that we represent all possible worldline perturbations with one solid line.

\section{Amplitudes}
\label{sec:amps}

\subsection{Compton amplitude}
As a first application of our Feynman rules we compute the Compton amplitude up to $S^4$. We then match to solutions of the Teukolsky equations~\cite{Bautista:2021wfy,Bautista:2022wjf} and fix Wilson coefficients in the effective action. For illustration purposes we go through the computations at $S^1$ from the minimal action, and then provide the full results at $S^4$. The Compton amplitude can be computed from the sum of three diagrams,
\begin{align}
\mathcal{A}^{\rm tree} =  \ \ \ \ 
\begin{tikzpicture}[rotate=180,scale=0.8,baseline={(0, 0.8cm)}]
	\draw[densely dotted] (0,0) -- (4,0);
	\draw[snake it,thick] (1,-2) -- (1,0);
	\draw[snake it,thick] (3,-2) -- (3,0);
	\draw[very thick] (1,0) -- (3,0);
	\draw[->] (2.3,-0.1) -- (1.7,-0.1);
	\node[] at (2,-0.4) {$\omega$};
	\node[] at (1,-2.3) {$k_2$};
	\node[] at (3,-2.3) {$k_1$};
\end{tikzpicture}
\ \ + \ \ 
\begin{tikzpicture}[rotate=180,scale=0.8,baseline={(0, 0.8cm)}]
	\draw[densely dotted] (0,0) -- (3,0);
	\draw[snake it,thick] (0.5,-2) -- (3/2,0);
	\draw[snake it,thick] (2.5,-2) -- (3/2,0);
	\node[] at (0.5,-2.3) {$k_2$};
	\node[] at (2.5,-2.3) {$k_1$};
\end{tikzpicture}
\ \ + \ \ 
\begin{tikzpicture}[rotate=180,scale=0.8,baseline={(0, 0.8cm)}]
	\draw[densely dotted] (0.5,0) -- (3.5,0);
	\draw[snake it,thick] (2,-1) -- (1,-2);
	\draw[snake it,thick] (3,-2) -- (2,-1);
	\draw[snake it,thick] (2,0) -- (2,-1);
	\node[] at (1,-2.3) {$k_2$};
	\node[] at (3,-2.3) {$k_1$};
\end{tikzpicture}
\end{align}
which we refer to as the $U$, $V$ and $Y$ diagrams respectively. Note that when expanding the exponential of the action in the path integral the $U$ diagram appears with a factor of $\frac{1}{2}$, but it combines with the crossed $U$ diagram. 
We take the external momenta to be incoming, and the flow of the internal energy $\omega$ in diagram $U$ is towards particle $2$, as indicated. This diagram has two delta functions at the vertices $\delta(k_1\md v -\omega)\delta(k_2\md v + \omega)$ which after integrating $\omega$ gives us the same delta function as in the other two diagrams, namely $\delta((k_1+k_2)\md v)$.

The $V$ diagram is trivial to compute, requiring only that we put the external gravitons on-shell. This means replacing $h_i^{\mu\nu}\to\e^\mu_i\e^\nu_i$ where $\e_i\md\e_i=\e_i\md k_i=k_i\md k_i=0$.
We also introduce the on-shell spin vector by replacing $S^{\mu\nu}\to \epsilon^{\mu\nu\rho\sigma}S_\rho v_\sigma$, such that the diagram is
\begin{align}
\mathcal{D}_V =& 
   \frac{i}{2} m  \e_1\md \e_2 v\md \e_1 v\md \e_2
   +\frac{1}{4}\e_1\md \e_2 v\md \e_2\epsilon^{vk_2 S \e_1}
   \nonumber \\ &\hspace{2cm}
   +\epsilon^{vS \e_1 \e_2}\Big(
   \frac{1}{4}v\md \e_1 q\md \e_2
   -\frac{1}{8}\e_1\md \e_2 v\md k_1\Big)
   +(1\leftrightarrow 2)
\ .
\end{align}
We made use of the shorthand notation $\epsilon^{vk_2 S \e_1}=\epsilon^{\mu\nu\rho\sigma}v_\mu k_{2\nu}S_\rho\e_{1\sigma}$ to keep the expression compact.
The $U$ diagram requires us to make use of the worldline propagators. This is easily implemented by effectively squaring the vertex in \eqref{eqn:Uvert} while assigning different energies to the placeholder worldline off-shell polarization tensors. We can then use the non-zero two-point functions in \eqref{eqn:propagators}. At zeroth order in spin we have the simple expression,
\begin{align}
\mathcal{D}_U\Big|_{S^0} = 
  \frac{i m}{2\omega} q\md\e_1 v\md\e_1 (v\md\e_2)^2
- \frac{i m}{2\omega} q\md\e_2 v\md\e_2 (v\md\e_1)^2
- \frac{i m}{4\omega^2} (v\md \e_1)^2 (v\md\e_2)^2 k_1\md q \ ,
\end{align}
where the worldline energy is $\omega = v\md k_1 = -v\md k_2$ and transferred momentum is $q=k_1+k_2$. At $S^1$ we find,
\begin{align}
\mathcal{D}_{U}\Big|_{S^1} =&
\frac{1}{4\omega}\Big(
\epsilon^{vk_1S\e_1}
\Big[
2\omega  v\md\e_2\e_1\md\e_2
- 2 v\md \e_1v\md \e_2 q\md\e_2
+  (v\md \e_2)^2 q\md \e_1
- \frac{1}{\omega} v\md \e_1 (v\md \e_2)^2 k_1\md q
\Big]
 \ \ \ \nonumber\\ &
+\epsilon^{vk_1S\e_2}v\md\e_1v\md\e_2q\md\e_1
+(1\leftrightarrow2)\Big) 
+ \frac{v\md \e_1 v\md\e_2}{4\omega}\Big(
\epsilon^{vS\e_1\e_2}  k_1\md q
+ \epsilon^{vk_1k_2S} \e_1\md\e_2 \Big) .
\end{align}
Finally, the $Y$ diagram is straightforward to put together using standard Feynman rules for gravity, its zero spin part is,
\begin{align}
\mathcal{D}_Y\Big|_{S^0} =&
\frac{im}{4 k_1\md k_2}\Big(
q\md \e_1 v\md \e_1 q\md \e_2 v\md \e_2
+2  \e_1\md \e_2 q\md \e_2 v\md \e_1 v\md k_1
-  k_1\md k_2\e_1\md \e_2 v\md \e_1 v\md \e_2
\nonumber \\ &\hspace{2cm}
- (v\md \e_1)^2 (q\md \e_2)^2
-\frac{1}{2}(\e_1\md \e_2)^2 (v\md k_1)^2
\Big) 
 + (1\leftrightarrow 2)
\end{align}
and at first order in spin,
\begin{align}
\mathcal{D}_Y\Big|_{S^1}=&
\frac{1}{4k_1.k_2}\Big(
\epsilon^{vqS\e_1}\big(
\e_1\md\e_2 q\md \e_2 v\md k_1 
- k_1\md k_2\e_1\md\e_2 v\md\e_2
-(q\md\e_2)^2 v\md\e_1 
+q\md\e_1 q\md\e_2 v\md\e_2 \big)
\nonumber \\ &\hspace{2cm}
+\epsilon^{vk_1qS}\big(
\e_1\md\e_2 q\md\e_2 v\md\e_1
-\frac{1}{2}(\e_1\md\e_2)^2 v\md k_1  \big)
+(1\leftrightarrow 2)
   \Big) \ .
\end{align}
The full covariant Compton amplitude is available in an ancillary file. There the Wilson coefficients in the action appear as free parameters, including $c_{RSK}$ and a tag keeping track of all acceleration terms.

As mentioned in the previous section the inclusion of the term \eqref{eqn:SK} in the action should be consistent with preservation of the SSC at $\mathcal{O}(S^2)$ as long as all acceleration terms are set to zero and one fixes $c_{RSK}\to 1$.
This was confirmed by comparing amplitudes computed with the acceleration terms (including from gauge fixing) to amplitudes computed with $c_{RSK}\to 1$ and all acceleration terms set to zero. Interestingly the two ways of computing the amplitude matched up to $S^4$, which indicates that additional terms from the MPD equation \eqref{eqn:mpd} do not appear in this order of the Compton amplitude. This is quite plausible as terms in the action that have more than one $K^\mu=S^{\mu\nu}p_\nu$ vector do not contribute to the Compton amplitude when the external states are fixed to obey the SSC.

Having the Compton amplitude at hand we can now compare it to the Teukolsky solutions~\cite{Bautista:2021wfy,Bautista:2022wjf} in order to fix our Wilson coefficients. Matching to either opposite or same helicity sectors is roughly the same, so we only show the details of the latter. In either case we find $C_{ES^n}=C_{BS^n} = 1$, together with three extra constraints on the six free coefficients at $\mathcal{O}(R^2)$. 
From the same helicity sector we can constrain differences, and from the opposite helicity matching we can constrain the sums of $E$- and $B$-type Wilson coefficients. Comparing the opposite helicity amplitudes to those in refs.~\cite{Bautista:2021wfy,Bautista:2022wjf} we find
\begin{align}\label{eqn:plusminusconstraints}
c_{E^2S^4} + c_{B^2S^4}= -1 \ , \ \ \ \ 
c'_{E^2S^4}+ c'_{B^2S^4} = 0 \ ,  \ \ \ \ 
c''_{E^2S^4} + c''_{B^2S^4}  = 0\ .
\end{align}
Next we plug in the positive-helicity polarization vectors $\e_1^{\mu} = \frac{\langle 2
|\sigma^\mu |1]}{\sqrt{2}\langle 12\rangle}$ and $\e_2^{\mu} = \frac{\langle 1|\sigma^\mu| 2]}{\sqrt{2}\langle 21\rangle}$ (we use the same massless spinor helicity conventions as ref.~\cite{Chiodaroli:2021eug}) as well as solve the constraints~\eqref{eqn:plusminusconstraints} for the $E$-type coefficients. We find the amplitude,
\begin{align}\label{eqn:pp}
\mathcal{A}^{++} =&  \mathcal{A}^{++}_{S=0}\Big(
1-\frac{q\md S}{m} + \frac{(q\md S)^2}{2m^2} - \frac{(q\md S)^3}{6m^3}  + \frac{(q\md S)^4}{24 m^4} -c_{B^2S^4}\mathcal{T} 
+c''_{B^2S^4}\frac{2\omega^2 q^2 S^4}{m^4} \\
&
+ c'_{B^2S^4}\frac{\omega^2 S^2(
	((8 \omega^2 - q^2) (q\md S)^2) + 
 q^2 ((\bar{q}\md S)^2 + 4 \omega^2 S^2) - 
 4 i\omega (q\md S) \epsilon^{q\bar{q}Sv}
)}{m^4 q^2}
\nonumber
\Big) \ ,
\end{align}
where $\bar{q} = k_2-k_1$, $q^2=|q|^2=-q\md q$,  the scalar amplitude is
\begin{align}
\mathcal{A}^{++}_{S=0} = 
\frac{im[12]^4}{\omega^2 q^2}  \ ,
\end{align}
and the extra term $\mathcal{T}$ is
\begin{align}
\mathcal{T} =& 
-\frac{4 \omega^4 S^2(((16 \omega^2 -  3 q^2) 
(q\md S)^2) 
  + q^2 ((\bar{q}\md S)^2 + 2 \omega^2 S^2) - 
   4 i \omega q\md S \epsilon^{q\bar{q}Sv})
   }{m^4 q^4}
   \nonumber\\
   &
   -\frac{ \omega^2 ((128 \omega^4 - 32 \omega^2 q^2 + 
      q^4) ({q}\md S)^4 +  
   2 q^2 (16 \omega^2 - q^2) (q\md S)^2 (\bar{q}\md S)^2 + 
   q^4 (\bar{q}\md S)^4)}{2 m^4 q^6}
   \nonumber\\
   &
   +\frac{4 i\omega^3 ((8 \omega^2 -q^2) (q\md S)^3 +  q^2 (q\md S) (\bar{q}\md S)^2) \epsilon^{q\bar{q}Sv}}{ m^4 q^6} \ .
\end{align}
In this amplitude we suppressed a factor of $\pi G$ from the coupling constant. The first line of \eqref{eqn:pp} clearly contains the exponential spin dependence of the Kerr amplitudes. The remaining terms are independent, and therefore their Wilson coefficients must be set to zero. The full set of constraints for Kerr scattering is collected in table \ref{tab:wc}
, and for completeness we give the resultant amplitude as well,
\begin{align}\label{eqn:ppfinal}
\mathcal{A}^{++} =&  \mathcal{A}^{++}_{S=0}\Big(
1-\frac{q\md S}{m} + \frac{(q\md S)^2}{2m^2} - \frac{(q\md S)^3}{6m^3}  + \frac{(q\md S)^4}{24 m^4}
\Big) \ .
\end{align}
\begin{table}[t]
\centering
\begin{tabular}{|c| c| c| c| c| c| c| c| c|}
\hline 
$c_{ES^2}$ & $c_{BS^3}$ & $c_{ES^4}$ & $c_{E^2S^4}$ & $c'_{E^2S^4}$ & $c''_{E^2S^4}$ & $c_{B^2S^4}$ & $c'_{B^2S^4}$ & $c''_{B^2S^4}$ \\
 \hline
 $1$ & $1$ & $1$ &$-1$ &$0$ &$0$ & $0$ & $0$ & $0$ \\
 \hline
\end{tabular}
\caption{Table of Wilson coefficients and their values for Kerr black hole scattering}
\label{tab:wc}
\end{table}

Our results are in agreement with the constraints obtained in ref.~\cite{Siemonsen:2019dsu}; first in order to match our conventions one has to redefine our spin vector and the $E$ and $B$ tensors by making use of the mass-shell constraint to modify our action such that $p^\mu/m\to \hat{p}^\mu+\ldots$. The mass-shell constraint gives 
\begin{equation}
m^2=p^2-\frac{c_{ES^2}E_{\mu\nu}S^\mu S^\nu}{m^2} +\ldots\ ,
\end{equation}
and applying this to the overall $m^{-4}$ in the $c_{ES^2}$ operator implies we find $(c_{E^2S^4}+1)$ instead of the old $c_{E^2S^4}$ as the coefficient of $(E_{\mu\nu} S^\mu S^\nu)^2$. Putting this together with the results in table \ref{tab:wc} we learn that if we used the unit-normalized $\hat{p}^\mu$ instead of $p^\mu/m$, all of our $\mathcal{O}(S^4R^2)$ Wilson coefficients go to zero, which is a solution to the constraints in equation (58) of ref.~\cite{Siemonsen:2019dsu}.

\subsection{One-loop amplitude}
The one-loop scattering for two spinning compact objects is computed along similar lines to the Compton amplitude.
We need to add to our path integral a second copy of a spinning particle action, such that the two spinning bodies can then interact gravitationally.
We label the particles and their respective masses $m_1$, $m_2$, spins $S_1$, $S_2$ and velocities $v_1$, $v_2$. The one-loop diagrams are roughly the same as those for the Compton amplitudes, but we add a second dotted line representing the background trajectory of particle number $2$. Our kinematic set up is captured by the diagram,
\begin{align}\label{eqn:loopblob}
\mathcal{A}^{\rm loop} =  \ \ \ \ 
\begin{tikzpicture}[rotate=180,scale=1,baseline={(0, 0.8cm)}]
	\draw[densely dotted] (0,0) -- (4,0);
	\draw[snake it,thick] (1,-2) -- (1,0);
	\draw[snake it,thick] (3,-2) -- (3,0);
	\draw[densely dotted] (0,-2) -- (4,-2);
	\fill[gray!50] (2,0) ellipse (1.3 and 0.4);
	\node[] at (3.5,-1) {$k$};
	\draw[->] (3.3,-1.4) -- (3.3,-0.6);
	\draw[->] (0.7,-1.4) -- (0.7,-0.6);
	\node[] at (0.2,-1) {$q-k$};
	\node[] at (-0.2,0) {$1$};
	\node[] at (-0.2,-2) {$2$};
\end{tikzpicture}
\end{align}
as well as its ``mirror image'' obtained by relabeling particles $1$ and $2$, and changing the sign of the transfer momentum $q$ (see for example~\cite{Kalin:2020mvi} for further details). In the diagram above the energy delta functions impose $v_2\md k = 0$ and $q\md v_2 = 0=q\md v_1$. Although this computation is classical we still have to compute loop integrals over the internal momentum $k$. 

Expanding the grey blob in~\eqref{eqn:loopblob}, the classical contribution to the one-loop amplitude is given by the diagrams,
\begin{align}
\mathcal{A}^{\rm loop} =  \ \ \ \ 
\begin{tikzpicture}[rotate=180,scale=0.8,baseline={(0, 0.8cm)}]
	\draw[densely dotted] (0,0) -- (4,0);
	\draw[snake it,thick] (1,-2) -- (1,0);
	\draw[snake it,thick] (3,-2) -- (3,0);
	\draw[very thick] (1,0) -- (3,0);
	\draw[densely dotted] (0,-2) -- (4,-2);
\end{tikzpicture}
\ \ + \ \ 
\begin{tikzpicture}[rotate=180,scale=0.8,baseline={(0, 0.8cm)}]
	\draw[densely dotted] (0,0) -- (3,0);
	\draw[snake it,thick] (0.5,-2) -- (3/2,0);
	\draw[snake it,thick] (2.5,-2) -- (3/2,0);
	\draw[densely dotted] (0,-2) -- (3,-2);
\end{tikzpicture}
\ \ + \ \ 
\begin{tikzpicture}[rotate=180,scale=0.8,baseline={(0, 0.8cm)}]
	\draw[densely dotted] (0.5,0) -- (3.5,0);
	\draw[snake it,thick] (2,-1) -- (1,-2);
	\draw[snake it,thick] (3,-2) -- (2,-1);
	\draw[snake it,thick] (2,0) -- (2,-1);
	\draw[densely dotted] (0.5,-2) -- (3.5,-2);
\end{tikzpicture}
\ .
\end{align}
The integral over $k$ contains three propagators, the massless $k^2$ and $(k-q)^2$ as well as the worldline energy propagator $\omega^{-2}=(v_1\md k)^{-2}$. Higher-order spin contributions also introduce a number of $k^\mu$ vectors in the numerators of the integrals. Here we find at most six loop momenta in the numerators,
\begin{equation}
I^{\mu_1\ldots\mu_6}(\nu_1,\nu_2,\nu_3) = \int \frac{d^4k}{(2\pi)^4}\frac{k^{\mu_1}\ldots k^{\mu_6}}{(k^2)^{\nu_1} ((k-q)^2)^{\nu_2} (k\md v_1)^{\nu_3}}\delta(v_2 \md k)\ ,
\end{equation}
where we restored the energy conserving delta function. 
The reduction of the tensor integrals to scalar integrals is done by making an ansatz for the tensors in terms of quantities transverse to $v_2$: $q^\mu$, $v_1^\mu-v_2^\mu\gamma$ and $\eta^{\mu\nu} - v_2^\mu v_2^\mu$, where $\gamma=v_1\md v_2$.
The scalar integrals have already been provided in the literature for generic propagators~\cite{Jakobsen:2021zvh}, thus our amplitude can be straightforwardly assembled. With the constraints on the Wilson coefficients presented in table \ref{tab:wc}, the resulting one-loop amplitude is presented below, with overall factors of $\pi^2 G^2$ from the coupling constant suppressed.
 It has been checked against the results in~\cite{Aoude:2022trd} with perfect agreement. The full one-loop amplitude in the ancillary file attached to this arXiv submission has all Wilson coefficients left free. 
\begingroup
\allowdisplaybreaks
\begin{align}
A\Big|_{S_1^0 S_2^0} =& 
	\frac{3  i  m_1 m_2 (m_1+m_2)  (1-5 \gamma^2)}{2 q}
\\
A\Big|_{S_1^1 S_2^0} =& 
	\frac{ m_2 (4 m_1+3 m_2)  \gamma (3-5 \gamma^2) \epsilon^{qS_1v_1v_2}}{2 q (-1+\gamma^2)}
\\
A\Big|_{S_1^2 S_2^0} =& 
	\frac{i  m_2 (8 m_2 + 60 m_2 \gamma^2 ( \gamma^2-1) + 
    m_1 (15 - 102 \gamma^2 + 95 \gamma^4)) ((q \md S_1)^2 + 
    q^2 S_1^2)}{16 m_1 q (1 - \gamma^2)} \nonumber \\
    &
    -\frac{i  m_2  q (9 m_1 + 2 m_2 - 6 (11 m_1 + 4 m_2) \gamma^2 + 
    5 (13 m_1 + 6 m_2) \gamma^4) (S_1 \md v_2)^2}{ 8 m_1 (1 - \gamma^2)^2}
\\
A\Big|_{S_1^1 S_2^1} =& 
	\frac{i  (m_1 + m_2)  (3 - 21 \gamma^2 + 
    20 \gamma^4) (q \md S_1 q \md S_2 + q^2 S_1 \md S_2)}{
 2 q (1 - \gamma^2)} \nonumber \\
    &
    +\frac{2 i  (m_1 + m_2)  q \gamma^3 (5 \gamma^2 -4) S_1 \md v_2 S_2 \md v_1}{(1 - \gamma^2)^2}
\\
A\Big|_{S_1^3 S_2^0} =& 
	-\frac{ m_2  \gamma (-5 m_1-2 m_2+(9 m_1+5 m_2) \gamma^2) ((q\md S_1)^2+q^2 S_1\md S_1) \epsilon^{qS_1v_1v_2}}{4 m_1^2 q (-1+\gamma^2)}\nonumber \\
    &
    -\frac{ m_2  q \gamma (-8 m_1+m_2+(16 m_1+5 m_2) \gamma^2) (S_1\md v_2)^2 \epsilon^{qS_1v_1v_2}}{4 m_1^2 (-1+\gamma^2)^2}
 \\
A\Big|_{S_1^2 S_2^1} =& 
 \frac{  q (4 m_2 (15 \gamma^2 -4 + 10 \gamma^4) + 
   m_1 (131 \gamma^4-12 - 11 \gamma^2 )) S_1 \md v_2 S_2 \md v_1 \epsilon^{q S_1 v_1 v_2}}{32 m_1 (\gamma^2 -1 )^2} \nonumber\\
  &
  +\frac{  \gamma (71 m_1 + 
    20 m_2 - (131 m_1 + 40 m_2) \gamma^2) (q \md S_1 q \md S_2 + 
    q^2 S_1 \md S_2) \epsilon^{q S_1 v_1 v_2}}{32 m_1 q (-1 + \gamma^2)}\nonumber\\
 &
  -\frac{ \gamma (60 m_2 (-1 + 2 \gamma^2) + 
    m_1 (-31 + 59 \gamma^2)) ((q \md S_1)^2 + q^2 S_1^2) \epsilon^{q S_2 v_1 v_2}}{32 m_1 q (\gamma^2-1 )}
    \nonumber\\
    &
    +\frac{  q \gamma (61 m_1 + 44 m_2 - 
   3 (43 m_1 + 40 m_2) \gamma^2) (S_1 \md v_2)^2 \epsilon^{q S_2 v_1 v_2}}{32 m_1 (\gamma^2-1)^2} 
\\
A\Big|_{S_1^2 S_2^2} =& 
    \frac{
    4 i q  m_1 (30 \gamma^2-3 - 29 \gamma^4) (S_1 \md v_2)^2 
    \big((q \md S_2)^2  + q^2 S_2^2\big)+(1\leftrightarrow 2)
       }{32 m_1 m_2 (1 - \gamma^2)^2}\nonumber
    \\&
    +\frac{
    i q m_2 (78 \gamma^2-5 - 81 \gamma^4) (S_1 \md v_2)^2 
    \big((q \md S_2)^2  + q^2 S_2^2\big)+(1\leftrightarrow 2)
       }{32 m_1 m_2 (1 - \gamma^2)^2}\nonumber
    \\&
    -\frac{i (m_1 + m_2)  q (8 - 85 \gamma^2 + 
    81 \gamma^4) \big((q \md S_2)^2 S_1 \md 
      S_1 + ((q \md S_1)^2 + q^2 S_1^2) S_2^2\big)}{
 32 m_1 m_2 (\gamma^2-1)}\nonumber
 \\&
 -\frac{i  (m_1 + m_2)  q (2 - 5 \gamma^2 + 7 \gamma^4) S_1 \md 
   S_2 (2 q \md S_1 q \md S_2 + q^2 S_1 \md S_2)}{16 m_1 m_2 (\gamma^2-1)}\nonumber
   \\&
   +\frac{i  (m_1 + m_2)  q \gamma (-9 + 10 \gamma^2 + 
   7 \gamma^4) (q \md S_1 q \md S_2 + q^2 S_1 \md S_2) S_1 \md v_2 S_2 \md 
  v_1}{8 m_1 m_2 (\gamma^2-1)^2}\nonumber
  \\
  &
  -\frac{i (m_1 + m_2)  q^{3} (5 - 79 \gamma^2 + 83 \gamma^4 + 7 \gamma^6) (S_1 \md 
    v_2)^2 (S_2 \md v_1)^2}{16 m_1 m_2 (\gamma^2-1)^3}\nonumber
\\
&
-\frac{i  (m_1 + m_2)  (12 - 95 \gamma^2 + 95 \gamma^4) (q \md 
    S_1)^2 (q \md S_2)^2}{32 m_1 m_2 q (\gamma^2-1)}
\\
A\Big|_{S_1^3 S_2^1} =& 
 \frac{4i m_2 (19 \gamma^2 -2 - 20 \gamma^4) \big((q \md S_1)^2 + q^2 S_1^2\big)
 (q \md S_1 q \md S_2 + q^2 S_1 \md S_2)}{
 48 m_1^2 q (\gamma^2-1)}\nonumber
 \\&
 +\frac{3i m_1 (37 \gamma^2 -5 - 36 \gamma^4) \big((q \md S_1)^2 + q^2 S_1^2\big)
 (q \md S_1 q \md S_2 + q^2 S_1 \md S_2)}{
 48 m_1^2 q (\gamma^2-1)}\nonumber
 \\&
    +\frac{i   q \big(6 m_1 (8 \gamma^2-1 - 8 \gamma^4) + 
    m_2 (1 + 13 \gamma^2 - 20 \gamma^4)\big) (q \md S_1 q \md S_2 + 
    q^2 S_1 \md S_2) (S_1 \md v_2)^2}{12 m_1^2 (\gamma^2-1)^2}\nonumber
    \\&
    -\frac{i q \gamma \big(2 m_2 (5 + 2 \gamma^2 - 10 \gamma^4) + 
    3 m_1 (1 + 6 \gamma^2 - 9 \gamma^4)\big) \big((q \md S_1)^2 + 
    q^2 S_1^2\big) S_1 \md v_2 S_2 \md v_1}{12 m_1^2 (\gamma^2-1 )^2}\nonumber
    \\&
    +\frac{i q^{3} \gamma \big(4 m_1 \gamma^2 (-5 + 6 \gamma^2) + 
   m_2 (-7 + \gamma^2 + 10 \gamma^4)\big) (S_1 \md v_2)^3 S_2 \md 
  v_1}{6 m_1^2 (\gamma^2-1)^3}
\\
A\Big|_{S_1^4 S_2^0} =& 
      \frac{i m_2 q \big(24 m_2 -25 m_1 +  2 (97 m_1 + 6 m_2) \gamma^2 \big) \big((q \md S_1)^2 + q^2 S_1^2\big) (S_1 \md v_2)^2}{96 m_1^3 (1 - \gamma^2)^2}\nonumber
      \\&
      -\frac{i m_2 q (193 m_1 + 60 m_2) \gamma^4 \big((q \md S_1)^2 + q^2 S_1^2\big) (S_1 \md v_2)^2}{96 m_1^3 (1 - \gamma^2)^2}\nonumber
      \\&
      +\frac{i m_2 q^{3} \big(8 m_1 (1 - 8 \gamma^2 + 8 \gamma^4) + 
    m_2 (-13 + 6 \gamma^2 + 15 \gamma^4)\big) (S_1 \md v_2)^4}
    {48 m_1^3 (1 - \gamma^2)^3}\nonumber
 \\&
 +\frac{i m_2 \big(24 m_2 \gamma^2 (-4 + 5 \gamma^2) + 
    m_1 (35 - 250 \gamma^2 + 239 \gamma^4)\big) \big((q \md S_1)^2 + 
    q^2 S_1^2\big)^2}{384 m_1^3 q (1 - \gamma^2)} \ .
\end{align}
\endgroup
We remind the reader that $q=|q|$ and $q^2=|q|^2 = -q\md q$. Additional parts of the amplitude containing higher powers of $S_2$ can be obtained by relabeling the components above.

\section{Conclusion}
\label{sec:con}

We have computed scattering amplitudes from a worldline EFT, employing the WQFT techniques. We were able to extend current computational tools up to fourth order in spin, and fix Wilson coefficients in the action by matching the scattering amplitudes to solutions of the Teukolsky equations, see table \ref{tab:wc} for those results. The six constraints we found on the Wilson coefficients are compatible with those identified previously in~\cite{Siemonsen:2019dsu} for certain linear combinations of our coefficients.  We additionally assembled the one-loop amplitude for the scattering of two spinning compact objects through $S^4$, and found our result is in agreement with the literature.
These computations relied on a new gauge-fixing condition for the point-particle EFT which allowed us to easily construct propagators and thus apply the formalism of the WQFT. We collected our amplitudes in an ancillary file, keeping all free parameters that appear in the action. 

For the gauge fixing to be valid one needs to know that the transformations in \eqref{eq:gaugetransforms} are not trivial, meaning that they are not proportional to the equations of motion. This requires checking that the constraints in the action are first class, which we show for the gauge-invariant action with generic non-minimal couplings in appendix \ref{sec:d}. The gauge fixing was also tested by keeping a free scalar parameter in front of the gauge-fixing condition and checking that the final results were independent of this parameter, these checks are explained in appendix \ref{sec:c}.
The invariance of the computations under the choice of this one scalar parameter in the action essentially amounts to checking a weakened Ward identity. We expect that since our gauge freedom here eliminates three degrees of freedom from the action we should be able to obtain three independent scalar Ward identities. Further investigation would be interesting and also useful because the Ward identities could help identify amplitudes that preserve the SSC even if they were guessed by a bootstrap procedure.

In this paper we worked up to quartic order in spin, but in principle there is no barrier to extend computations to higher orders. It would therefore be interesting to extend our Feynman rules and use them to study higher-spin effects in radiation~\cite{DeAngelis:2023lvf,brandhuber2023resummed,aoude2023leadingorder}, as well as higher loop corrections to scattering amplitudes and potentials. Since computations involving gravity Feynman rules can become quite cumbersome, it would be useful to develop a double-copy prescription~\cite{Shi:2021qsb} for the classical spinning EFT. 
The techniques developed here could also be extended to include non-conservative effects. These can be added in the worldline equations of motion for example~\cite{Saketh:2022xjb}, and such equations of motion could still in principle be solved perturbatively using Feynman rules similar to the ones we developed here.
Inclusion of non-conservative effects would allow for a full matching to Kerr scattering results beyond fourth order in spin~\cite{Bautista:2021wfy,Bautista:2022wjf}, but even without them it could be possible to identify worldline actions that match conservative sectors of Kerr Compton amplitudes beyond $S^4$. Such actions could shed light on the guiding principles for constraining effective actions for Kerr black holes at higher orders in curvature. We leave these topics for future research.

\section*{Acknowledgments}
I would like to thank Francesco Alessio, Zvi Bern, Marco Chiodaroli, Gustav Jakobsen, Henrik Johansson, Jung-Wook Kim, Gustav Mogull, Paolo Pichini, Jan Plefka, and Jan Steinhoff for useful discussions.
I am also grateful to Justin Vines, Henrik Johansson, and Zhewei Yin for illuminating discussions and useful feedback on an earlier draft of this work.
Finally I am particularly grateful to Lucile Cangemi and Kays Haddad for their helpful feedback and for providing materials for comparisons.
I would like to acknowledge the use of FORM~\cite{Vermaseren:2000nd} in some of these computations.
This research was supported in part by the Knut and Alice Wallenberg Foundation under grants KAW 2018.0116 ({\it From Scattering Amplitudes to Gravitational Waves}) and KAW 2018.0162 ({\it Exploring a Web of Gravitational Theories through Gauge-Theory Methods}). 

\appendix

\section{The worldline action}
\label{sec:A}
The full worldline action up to $S^4$ and before gauge fixing is,
\begin{align}
S =& -\int d\tau \Big(
	p_\mu \dot{x}^\mu 
	+ \frac{1}{2} S_{ab}\Omega^{ab}
	+ \frac{1}{p^2}\frac{D p_\mu}{d\tau} S^{\mu\nu}p_\nu 
	-\frac{\ell}{2}(p^2-m^2)
	-\ell_ aS^{a b}(\hat{p}_b + \Lambda_{b 0}) \nonumber\\
	&\hspace{1.6cm}
	+\frac{\ell c_{ES^2}}{2m^2}E_{\mu\nu}S^\mu S^\nu
	-\frac{\ell c_{RSK}}{2m^2} R_{\mu\nu\rho\sigma}S^{\mu\nu}K^\rho p^\sigma
	- \frac{\ell c_{BS^3}}{6m^3}D_\mu B_{\nu\rho}S^\mu S^\nu S^\rho \nonumber\\
	&\hspace{1.6cm}
	-\frac{\ell c_{ES^4}}{24m^4}D_\mu D_\nu E_{\rho\sigma}S^\mu S^\nu S^\rho S^\sigma \\
	&\hspace{1.6cm}
	+\frac{\ell c_{E^2S^4}}{ m^6}(E_{\mu\nu}S^\mu S^\nu)^2
	+ \frac{\ell c_{B^2S^4}}{m^6} (B_{\mu\nu}S^\mu S^\nu)^2
	+\frac{\ell c'_{E^2S^4}}{m^6} S^2 S^\mu E_{\mu\nu}E^\nu{}_{\rho}S^\rho\nonumber
	\\
	&\hspace{1.6cm}
	+\frac{\ell c'_{B^2S^4}}{m^6} S^2 S^\mu B_{\mu\nu}B^\nu{}_{\rho}S^\rho
	+\frac{\ell c''_{E^2S^4}}{m^6} S^4  E^2
	+\frac{\ell c''_{B^2S^4}}{m^6} S^4  B^2
	\Big)\nonumber
\end{align}
where $D/d\tau$ is the covariant derivative along the worldline, e.g. $Dp_\mu/d\tau=\dot{p}_\mu - \Gamma^\rho_{\mu\nu}p_\rho\dot{x}^\nu$. For completeness, the angular velocity is defined by,
\begin{align}
\Omega^{ab} =& \Lambda_I^a\frac{D}{d\tau}\Lambda^{bI} = \Lambda_I^a \dot{\Lambda}^{bI} - \omega_{\mu}{}^{ab} \ ,
\end{align}
where $\omega_\mu^{ab}=e^{a}_\nu\partial_\mu e^{\nu b} + \Gamma_{\mu\nu}^\rho e^a_\mu e^{b\nu}$ is the spin connection for the flat corotating frame of the initial state.

\section{Composite spin tensor}
\label{sec:B}
Our starting point was a worldline with six degrees of freedom contained in the $\Lambda_I^a$, and six conjugate degrees of freedom contained in $S_{ab}$. It is natural to ask if instead of $S_{ab}$ we could have introduced a matrix $\chi_a^I$ conjugate to the $\Lambda_I^a$. It turns out that we can via the action,
\begin{align}
S_{\textrm{comp. spin}} = -\int d\tau \Big(
	\chi_a^I\dot{\Lambda}^a_I  - \ell_{IJ}(\Lambda_a^I\Lambda^{aJ}-\eta^{IJ}) 
\Big)\ .
\end{align}
The constraint $(\Lambda_a^I\Lambda^{aJ}-\eta^{IJ})$ generates a gauge symmetry,
\begin{align}
	\label{eqn:chigauge}
	\delta_{\epsilon} \chi_a^I = -2\epsilon_{IJ}\Lambda_a^J \ ,
\end{align}
which reduces the number of physical degrees of freedom in $\chi$ to six.
It is straightforward to see that the composite spin tensor defined by
\begin{align}
S_{ab} = \Lambda_a^I\chi_{bI} - \lambda_b^I\chi_{aI} \ ,
\end{align}
is gauge invariant under the gauge transformations \eqref{eqn:chigauge}, and that the kinetic terms with $S_{ab}$ or with $\chi_a^I$ are equal
\begin{align}
\frac{1}{2}S_{ab}\Omega^{ab} = \chi_a^I \frac{D\Lambda^a_I}{d\tau} \ .
\end{align}

\section{$R_\xi$ gauge fixing}
\label{sec:c}
The gauge fixing condition was relaxed to,
\begin{align}
\ell_\mu = \frac{\xi}{p}\frac{Dp_\mu}{d\tau} \ ,
\end{align}
where $\xi$ was left as a free parameter throughout the computations. With this choice of gauge fixing the two point functions become,
\begin{align}\label{eqn:generalpropagators}
	\langle z^\mu(-\omega) z^\nu(\omega)\rangle =& -i\frac{1}{m\omega^2}\eta^{\mu\nu} - \alpha\frac{1}{m^2 \omega} {S}^{\mu\nu}\ ,  \\
	\langle p^\mu(-\omega) z^\nu(\omega)\rangle =& -\frac{1}{\omega}\eta^{\mu\nu}\ ,  \\
	\langle s_{\mu\nu}(-\omega)s_{\rho\sigma}(\omega) \rangle =&  -\frac{4}{\omega} \eta_{\nu][\sigma}{S}_{\rho][\mu}\ , \\
	\langle s_{\mu\nu}(-\omega)\lambda_{\rho\sigma}(\omega)\rangle =& \frac{2}{\omega}\eta_{\mu[\rho}\eta_{\sigma]\nu}\ ,  \\
	\langle \lambda^{\mu\nu}(-\omega)z_\rho(\omega)\rangle =& -\alpha(2\xi-1)\frac{2}{m\omega}v^{[\mu}\delta^{\nu]}_\rho \ ,  \\
	\langle z^\mu(-\omega)s^{\nu\rho}(\omega)\rangle =& \alpha (1-\xi)\frac{2}{\omega}S^{\mu[\nu}v^{\rho]} \ .
\end{align}
In addition to the $\xi$ parameter we also introduced $\alpha$, which multiplies everything to do with gauge fixing and acceleration terms. The SSC is preserved by setting $\alpha\to 1$, and $\xi$ can be arbitrary. We also checked that the SSC is preserved when setting $\alpha\to 0$ and $c_{RSK}\to 1$. If one instead sends $\alpha\to 0$ and keeps $c_{RSK}$ free the SSC will be broken. In the case when the SSC is kept, we have verified that the two ways of computing the Compton amplitude give identical results, and also that the $\alpha\to 1$ method is independent of $\xi$. These checks were done through fourth order in spin.

\section{First-class constraints}
\label{sec:d}
Here we show that the constraint $S\md (\hat{p}+\Lambda_0)$ is first class, whenever the non-minimal interactions preserve the gauge symmetry. This is equivalent to checking if the constraints are preserved in time on the constraint surface. 
We rewrite the action as
\begin{align}
S =& -\int d\tau \Big(
	p_\mu \dot{x}^\mu 
	+ \frac{1}{2} S_{ab}\Omega^{ab}
	+ \frac{D \hat{p}_\mu}{d\tau} S^{\mu\nu}\hat{p}_\nu 
	-\frac{\ell}{2}\big(p^2-\mathcal{N}\big)
	-\ell_ aS^{a b}(\hat{p}_b + \Lambda_{b 0}) 
	\Big)\ ,
\end{align}
where the dynamical mass squarred $\mathcal{N}=m^2 +\ldots$ was introduced and includes all of the non-minimal interactions. 
We need only use the two equations of motion,
\begin{align}
\Omega^{\mu\nu} +2\frac{D \hat{p}^{[\mu}}{d\tau}\hat{p}^{\nu]} 
- 2 \ell^{[\mu}\big(\hat{p}^{\nu]} + \Lambda_0^{\nu]}\big) +\ell \mathcal{N}^{\mu\nu}&=0 \ , \\
\frac{D}{d\tau}S_{\mu\nu} -2S_{[\mu|\rho}\Omega^\rho{}_{\nu]} - 2\ell_\rho S^{\rho}{}_{[\mu}\Lambda_{\nu]0} &= 0 \ ,
\end{align}
where we introduced $\mathcal{N}^{\mu\nu}=\frac{\partial\mathcal{N}}{\partial S_{\mu\nu}}$. Gauge invariance required us to write $\mathcal{N}$ only in terms of the spin vector, or spin tensor contracted with projectors transverse to $p^\mu$, therefore we have $\mathcal{N}_{\mu\nu}p^\nu = 0$.
Using the equation of motion for $S_{\mu\nu}$ we find,
\begin{align}
\frac{D}{d\tau} S^{\mu\nu}(\hat{p}_\nu + \Lambda_{\nu 0}) &=
\big( 2S_{[\mu|\rho}\Omega^{\rho}{}_{\nu} + 2 \ell_\rho S^{\rho}{}_{[\mu}\Lambda_{\nu]0}\big)\big(\hat{p}^\nu + \Lambda^\nu_0\big) + S^{\mu\nu}\Big(\frac{D\hat{p}_\nu}{d\tau} - \Omega_\nu{}^\rho\Lambda_{\rho 0}\Big) \\
&\approx
\big( S_{\mu\rho}\Omega^{\rho}{}_{\nu} +  \ell_\rho S^{\rho}{}_{\mu}\Lambda_{\nu0}\big)\big(\hat{p}^\nu + \Lambda^\nu_0\big) + S^{\mu\nu}\Big(\frac{D\hat{p}_\nu}{d\tau} - \Omega_\nu{}^\rho\Lambda_{\rho 0}\Big)  \\
&=S^{\mu\nu}\frac{D\hat{p}_\nu}{d\tau} + S_{\mu\rho}\Omega^{\rho}{}_\nu \hat{p}^\nu + \ell_\rho S^{\rho\mu}(\Lambda_0\md \hat{p} + 1) \ , 
\end{align}
where $\approx$ means equality on the constraint surface. Next we substitute $\Omega_{\mu\nu}$, and working on the constraint surface as above, we find 
\begin{align}
\frac{D}{d\tau} S^{\mu\nu}(\hat{p}_\nu + \Lambda_{\nu 0}) &\approx S_{\mu\rho} \hat{p}^\rho \frac{D\hat{p}^\nu}{d\tau}\hat{p}_\nu \ ,
\end{align}
which is zero due to the fact that $\hat{p}$ is unit normalized.

\bibliographystyle{JHEP}
\bibliography{bib}
\end{document}